# Observation of image pair creation and annihilation from superluminal scattering sources

M. Clerici[1,2], G.C. Spalding[3], R. Warburton[1], A. Lyons[1], C. Aniculaesei[1], J.M. Richards[3], J. Leach[1], R. Henderson[4], D. Faccio[1,*]

[1] School of Engineering and Physical Sciences, SUPA, Heriot-Watt University, Edinburgh EH14 4AS, UK.
[2] School of Engineering, University of Glasgow, Glasgow G12 8LT, UK.
[3] Dept. of Physics, Illinois Wesleyan University, Bloomington, IL 61701, USA.
[4] Institute for Micro and Nano Systems, University of Edinburgh, Edinburgh EH9 3FF, UK.
* Correspondence to: d.faccio@hw.ac.uk.

**The invariance of the speed of light implies a series of consequences related to our perception of simultaneity and of time itself. Whilst these consequences are experimentally well studied for subluminal speeds, the kinematics of superluminal motion lack direct evidence. Using high temporal resolution imaging techniques, we demonstrate that if a source approaches an observer at superluminal speeds, the temporal ordering of events is inverted and its image appears to propagate backwards. If the source changes its speed, crossing the interface between sub- and super-luminal propagation, we observe image pair annihilation and creation. These results show that it is not possible to unambiguously determine the kinematics of an event from imaging and time-resolved measurements alone.**

Displaying prescient intuition, Lord Rayleigh noted that a supersonic source of sound waves could give rise to time reversal of the perceived sound by a stationary observer. For the specific one-dimensional case in which the source moves at exactly twice the speed of sound, ``sounds previously excited would be gradually overtaken and heard in reverse of natural order…the observer would hear a musical piece in correct time and tune, but *backwards*.'' [1]. Unfortunately, any attempt to actually play out such an experiment is faced with wave attenuation over the huge distances (~1 km) covered by a supersonic source whilst emitting just three seconds of music. However, the reasoning followed by Lord Rayleigh relies solely on the fact that the wave speed is finite and independent of the speed of the emitter. The same result therefore also holds true for light waves.

Contrary to typical expectations, it is possible to create a superluminal source of light, where we use the term ``source'' in a very broad sense. Consider, for example, a wavefront impinging on a flat surface such as a wall: the intersection point of the wavefront with the wall moves at a speed $v = c/\sin\theta$, where $c$ is the speed of light in vacuum and $\theta$ is the angle made between the normal to the wall surface and the wavevector. Therefore, $v > c$ for all wavefront propagation angles. Moreover, this intersection point will, in general, always be visible due to scattering from the wall surface. Hence, although there is no physical source of light moving at $v > c$, we nevertheless have a superluminal ``scattering source'' that can be used to study and observe the *kinematics* of superluminal phenomena.

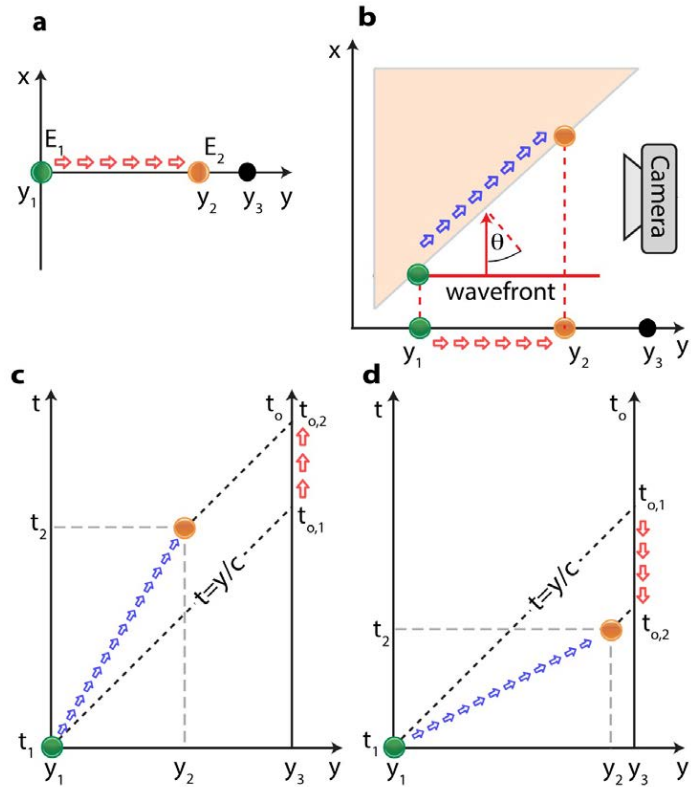

**Figure 1. a** illustrates the (1+1)D example described in the text. **b** shows the extension of **a** to a (2+1)D case that exemplifies the experimental layout. The motion of the scattering source towards the observer (red arrows) can be either superluminal or subluminal. **c** Minkowski diagram for two sequential events: because *this case has* $v < c$ time-ordering is preserved. **d** Minkowski diagram for two (causally disconnected) events where $v > c$: time-ordering is inverted.

Superluminal sources or more precisely, sources with a group velocity exceeding the vacuum speed of light $c$, were given a precise description by Brillouin [2,3] and then observed in a number of different optical arrangements, for example in 'fast-light' media [4], in the propagation of Bessel beams [5-9], in Lyot filters [10] and from scattering surfaces [11]. Whilst it is now accepted that superluminal group velocity does not contradict the theory of special relativity since the information speed is always limited by $c$ (see *e.g.* [12]), to our knowledge time-ordering or, in general, the kinematics associated with superluminal speeds has not yet been experimentally addressed, and there has been no prior demonstration of the image pair creation and annihilation shown in this work.

Here we present a series of experiments that rely on ultrafast imaging techniques, which illustrate various kinematical phenomena, including time reversal and image pair creation and annihilation at transitions from sub to superluminal propagation.

For illustration purposes, we first consider the simplest (1+1)D [$x+t$] situation sketched in Fig.1a where we consider two events, $E_1(y_1, t_1)$ and $E_2(y_2, t_2)$, taking place at two separate positions ($y_1$, $y_2$) and times ($t_1$, $t_2$), associated with a moving source. We also consider an observer with a camera in a fixed reference frame identified as the laboratory frame and at a position $y_3$. Whilst the original time delay between $E_2$ and $E_1$ is ($\Delta t = t_2 - t_1$), the time difference recorded by the observer (at position $y_3$) is obtained considering that information of these events travels to the camera at the speed of light $c$:

$$\Delta t_{observer} = \left(t_2 + \frac{y_3-y_2}{c}\right) - \left(t_1 + \frac{y_3-y_1}{c}\right) = \Delta t \left(1 - \frac{v}{c}\right), \quad \text{Eq. 1}$$

where $v$ is the speed of the source along the y-direction. For $v < c$ the observer will perceive a reduced time delay but the time-ordering of the events is preserved. However, if $v > c$ the time-ordering will be inverted as $\Delta t_{observer} < 0$. In other words, if the observer is trying to record the image of a superluminal object then they will not be able to tell from the time-resolved video data alone whether the source is approaching or moving from away from them.

The geometry we investigate in our experiments (Fig. 1b) is that of a superluminal spot (blue arrows) created by a pulsed plane wave as shown (with a red line), which is itself propagating at the speed of light and impinging on a tilted screen. As already noted, this spot propagates at speeds relative to the screen that are *superluminal* regardless of the screen angle. However, the component of the velocity of the spot along the direction toward the observer (here, along the y-direction) will, in general, depend on the screen angle. This latter velocity can therefore be experimentally tuned from sub- to super-luminal by simply tuning the screen inclination angle.

The perceived temporal inversion of events relies only on two ingredients: the wave speed should be finite and independent of the emitter speed and the emitter should be moving faster than the wave speed *in the direction of the observer*. The generality of these conditions can be seen by plotting the kinematics in the relative Minkowski space-time diagrams for the subluminal (Fig. 1c) and superluminal cases (Fig. 1d). In the subluminal case, the worldline of the emitter (blue arrows) lies above the lightline ($t = y/c$) and the measured arrival times of these emanations, for a stationary observer at $y_3$, retain proper time-ordering (indicated by the red arrows). Conversely, a superluminal emitter's worldline lies in the region below the lightline. Geometrical construction of the stationary observer's measurements of the same events shows that these must be characterized by a time-ordering inversion (indicated by the downward orientation of the red arrows in Fig. 1d).

We underline that although superluminal motion is involved here, there is no superluminal transfer of information since the scattering events at distinct regions of the screen are not causally connected (they belong to physically distinct regions of the incoming wavefront). Moreover, we do not need to consider relativistic effects or Doppler shifts since there are no dipole emitters that are actually moving.

For the geometry established by Fig. 1b, the component of the spot velocity in the y-direction of the observer is simply given by $v = c \cdot \cot\theta$. Therefore, the component of the spot velocity in the direction of the observer is superluminal for $0 < \theta < \pi/4$ and subluminal for $\pi/4 < \theta < \pi/2$. Following the considerations from Eq. 1, the observer will record an inverted time-order of the events for the former case. Straightforward generalization of this argument reveals that time-ordering inversion results whenever the angle of detection is greater than the angle of incidence. Furthermore, the inverted time ordering also modifies the observer's perception of the speed of the scattering source along the $\hat{x}$ direction. Indeed, the recorded speed along $\hat{x}$ is:

$$v_x^o = \frac{dx}{dt_o(x)} = \frac{c}{1-\cot\theta} \quad \text{Eq. 2}$$

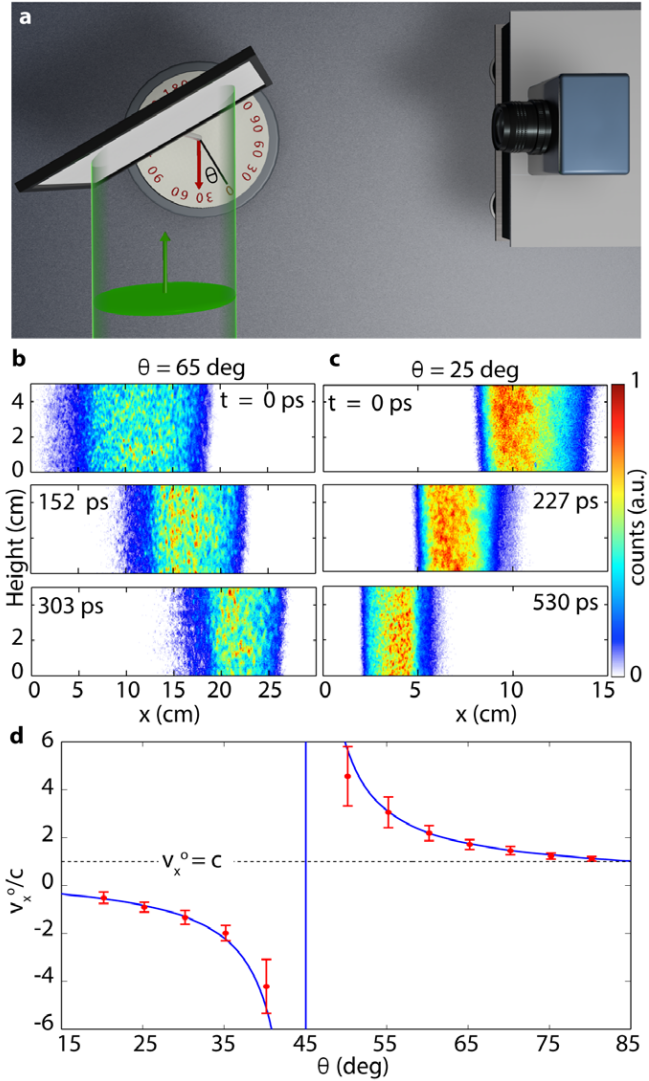

**Figure 2 a** a sketch of the experiment. A plane wavefront (green) impinges on a tilted screen and the scattered radiation is recorded at 90 deg with a time-resolving iCCD camera. Changing the angle $\theta$ between the input wave and the screen allows a change in the scattering source velocity component along the camera/observer direction. **b** shows three snapshots acquired by the camera at three different times for an incident angle ($\theta$ =65 deg) such that the scattering speed towards the camera is subluminal. In this case the time order is maintained and the perceived source moves from left to right. **c** for $\theta$ = 25 deg the source velocity towards the detector is superluminal and event time-ordering is reversed, *i.e.* the same wavefront measured in **b** is now seen as propagating in the opposite direction, from right to left. **d** shows the measured measured speed along the *x* direction (red dots) compared with the theoretical prediction (blue curve).

where the time $t_o(x)$ is the arrival time of the signal on the *x* position of the detector. Therefore, for $0 < \theta < \pi/4$ the perceived speed along the $\hat{x}$ direction has an opposite sign with respect to the real speed. In such circumstances, even a detector with sufficient resolution to track the events will not be able to distinguish between a source moving from left to right at superluminal speed from one moving in the opposite direction at subluminal speed.

The experimental setup is shown in Fig. 2a: a time resolving camera is used to image in-plane scattering from the wavefront generated by a diffused 130-fs, laser pulse impinging on an inclined surface with 50x50 cm$^2$ area. The camera is a time-resolving intensified charge-coupled device (iCCD) that acquires 520×688 pixel images with a 200-ps temporal gate timed to the laser pulse. Enhanced temporal resolution is achieved by means of a delay generator with 10-ps temporal resolution (LaVision, PicoStar).

Figures 2b show a temporal sequence of images taken from the full video for the subluminal case ($\theta$ = 65 deg, i.e. $v = c \cdot \cot\theta = 0.46c$). We see the wavefront propagating across the screen from left to right, i.e. with the correct temporal ordering of events. Figures 2c show the same sequence but now for the case in which the scattering event has superluminal speed in the direction of the camera ($\theta$ = 25 deg, i.e. $v = c \cdot \cot\theta = 2.14c$). The wavefront is now seen to propagate in the opposite direction so temporal ordering is clearly inverted. In Fig. 2d we compare the measured speed $v_x^o$ with the prediction of Eq. 2 while systematically increasing the incident angle $\theta$ – the results show very good agreement with the predictions.

## *Image pair creation and annihilation*

So far we have considered the simple case of a flat, tilted scattering screen leading to uniform motion of the source. Interesting effects arise when the source has non-uniform motion, in particular with a sub to superluminal transition (or vice-versa). Such a situation is obtained by adequately curving the scattering surface. Without loss of generality we consider the case of a scattering screen described by the function $S(x) = x^2$. Following the very same arguments reported above we find:

$$v_x^o = \frac{c}{1-2x}, \quad \text{(a)}$$
$$\frac{dt_o}{dt} = \frac{1-2x}{c}. \quad \text{(b)}$$

Eq. 3

Clearly, the perceived speed along the $x$ direction is positive for $x < 0.5$ and negative for $x > 0.5$, resulting in two images moving in opposite directions along the $x$ axis [from left to right for $x < 0.5$ and from right to left for $x > 0$, Eq. 3(a)]. Correspondingly, the temporal axis at the observation plane is be reversed for $x > 0.5$ [Eq. 3(b)]. As sketched in Fig. 3, the observer will therefore see two stripes of light that move towards each other and disappear at $x = 0.5$, a process that we refer to as image pair annihilation. Changing the sign of the curvature of the surface $S(x)$ will result in the opposite process: taking for example $S(x) = -(x-1)^2$, we have $v_x^o = c(2x-1)^{-1}$, and the observer will perceive the light wave scattering on the surface as an image pair creation, originating from $x = 0.5$. A similar prediction was recently made by R. J. Nemiroff albeit in an astronomical setting, e.g. where the curved surface is represented by the edge of the Moon [13].

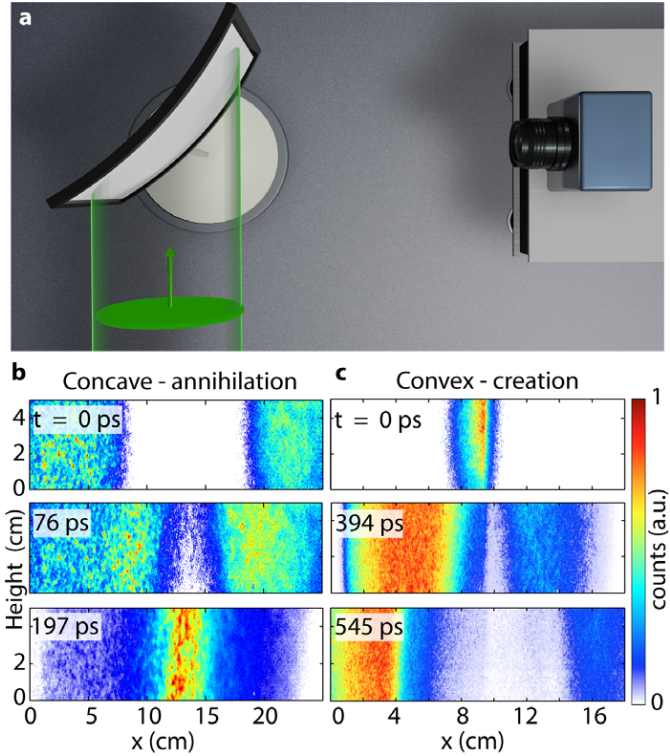

**Figure 3**. Image pair annihilation and creation: **a** layout of the experiment. **b** shows three snapshots acquired by the camera at three different times for concave screen, resulting in the annihilation of image pairs. **c** shows three acquisitions for a convex screen, resulting in the creation of image pairs.

In Figure 3 we show an example of an experiment performed using a curved scattering screen (see Fig. 3a) and illustrating the annihilation and generation of image pairs. For a properly chosen concave shape the camera records the annihilation of an image pair, as shown in the three acquisitions reported in Fig. 3b. Similarly, a convex screen results in the creation of an image pair, as shown in Fig. 3c. We underline that at any given time the propagating wavefront has one and only one intersection point (the scattering source) along any horizontal line on the screen: the observed image splitting is therefore truly a result of the transition between sub and superluminal propagation (see the *Appendix*, below).

## *Discussion*

The crucial ingredient required in these experiments is that the geometry allows propagation in the direction of the observer that is faster than the free-space wave propagation speed. Similar image pair effects were predicted in an astronomical context where they might however be somewhat harder to observe [13]. We note here that these effects are not to be confused with the *apparent* superluminal motion of astrophysical radio sources [14] that are due to movement perpendicular to or at an angle with respect to the observer. Here, the observed effects are due to a *real* superluminality in the direction of the observer.

Similarly to light, the propagation of sound waves or mechanical vibrations may give rise to temporal inversion. Aside from the predictions by Lord Rayleigh, a further example worth considering would be the scattering of seismic waves from an inclined geological surface. The detection of scattered seismic waves is commonly used to determine the composition of the inner layers of the Earth's structure yet it is clear from the considerations above that a temporally resolved measurement could give rise to an apparently inverted geophysical structure. Notably, the conclusion is that it is not possible to unambiguously determine the ``true'' kinematics of an event by relying solely upon imaging and time-resolved measurements. This ambiguity could be removed, for example, by acquiring additional information either regarding the exact conformation of the scattering surfaces, speed of the source or the source coordinate along the direction of observation. Such considerations will play a crucial role in emerging time-resolved imaging technologies that, indeed, rely on detecting scattered light from surfaces [15-20].

Superluminal motion and its implications have also been widely discussed in a number of contexts such as tachyonic particles or superluminal tunneling [21-25]. In particular, the precise form of the equivalent Lorentz transform for superluminal motion has been widely debated [26] with some open problems still remaining unresolved [27]. Experiments such as those shown here could be adapted to provide experimental grounding for the assumptions that underlie such theoretical models.

## *Appendix*

Here we provide additional theoretical analysis and we consider a more general situation for arbitrary observation and wavefront incidence angle. We also report experimental observations performed for various incident and observation angles, acquired with a different detector (a single photon avalanche photodetector array – SPAD) with respect to that employed in the main text. Finally, we expand on the origin of the image pair creation and annihilation effects.

**General relation for measured velocities**

We consider the more general situation depicted in Fig. S1, where an incident plane wave with wavevector $k$ scatters from a flat screen tilted in such a way that its normal forms an angle $\theta$ with $k$. We first analyse the case of in-plane scattering, where the camera is placed in such a way that the observation direction forms an angle $\phi$ with the normal to the scattering surface. We wish to evaluate the transformation of the time coordinate into the measured time and the consequent transformation between the speed along the $x'$ direction and the *measured* speed along $x'$. To this end, we start by noting that in the $(x,y)$ reference frame,

$$x(t) = ct$$
$$y(t) = \cot(\theta) x(t)$$

We can hence evaluate the spatial coordinates of the scattering source in a new reference frame $(x', y')$, as shown in Fig. S1. The angle between the incident wavevector, which defines our $x$-axis, and the path from screen to detector, which defines the $y'$-axis, is $(\theta + \phi)$. Thus, the angle between the $x$- and $x'$-axes is $(\theta + \phi - \pi/2)$, and so the coordinate transformation is given by:

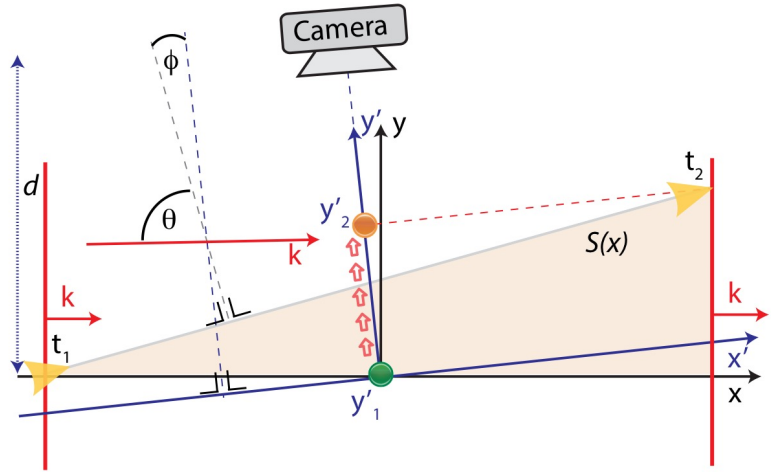

**Figure S1.** Layout of the situation described in the text, for an angle of observation $\phi$ independent from the angle of incidence $\theta$.

$$x'(x) = \cos\left[(\theta+\phi)-\frac{\pi}{2}\right]x - \sin\left[(\theta+\phi)-\frac{\pi}{2}\right]y(x) = x\cos(\phi)\csc(\theta)$$

$$y'(x) = \sin\left[(\theta+\phi)-\frac{\pi}{2}\right]x + \cos\left[(\theta+\phi)-\frac{\pi}{2}\right]y(x) = x\csc(\theta)\sin(\phi)$$

Which leads to:

$$x'(t) = ct\cos(\phi)\csc(\theta)$$
$$y'(t) = ct\csc(\theta)\sin(\phi)$$

In this case, the speed along $x'$ and $y'$ can be evaluated as

$$v_{x'} = \frac{dx'}{dt} = c\cos(\phi)\csc(\theta)$$
$$v_{y'} = \frac{dy'}{dt} = c\csc(\theta)\sin(\phi)$$

Eq. S1

In order to find the relation linking the measured time interval ($t_o$) with the original one ($t$) and the measured speed ($v_{x'}^0$) to the actual spot speed relative to the screen ($v_{x'}$), we consider that $t_o$ is the arrival time of the information at the plane of the detector and is obtained by summing two components: $t_A$, the interval of time over which the scattering source moves from the origin to the point $y'$ and $t_B$, the time required to the scattered light to propagate in free space from $y'$ to the detector. These read:

$$t_A = t$$
$$t_B = \frac{d-y'(t)}{c}$$

so that:

$$t_o = t + \frac{d - ct\csc(\theta)\sin(\phi)}{c}$$

and therefore the relation between the measured and the real time is:

$$\frac{dt_o}{dt} = 1 - \csc(\theta)\sin(\phi) = 1 - \frac{v_{y'}}{c}$$

Eq. S2

and the measured velocity along $x'$ reads:

$$v_{x'}^o = \frac{dx'(t_p)}{dt_p} = \frac{dx'(t)}{dt}\frac{dt}{dt_o} = \frac{c\cos(\phi)}{\sin(\theta) - \sin(\phi)}$$

Eq. S3

We note that these equations correctly reduce to those considered in the main text. Indeed, for $\phi = \pi/2 - \theta$, Eq. S2 reduces to:

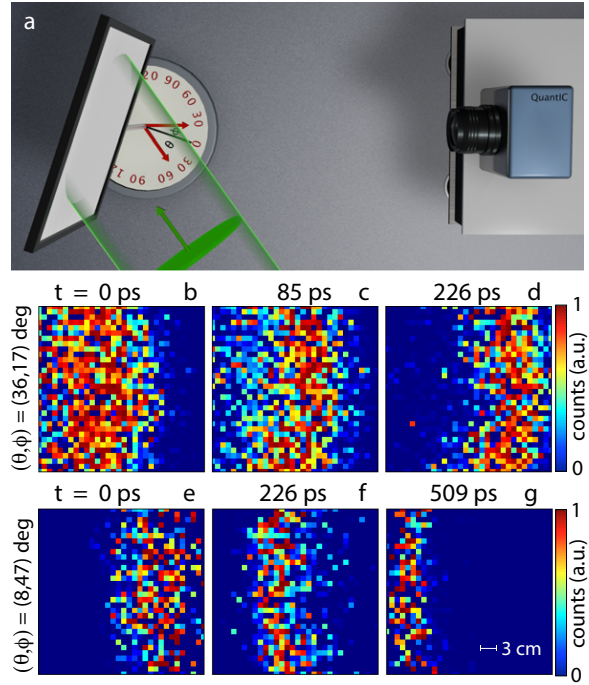

**Figure S2. a** shows a sketch of the experimental setup employed to perform the time-resolved imaging for the conditions reported above, with a SPAD camera detector. **b-c** show the images recorded for three different instant of time for subluminal propagation, whereas **e-g** show the same for a superluminal source.

$$\frac{dt_o}{dt} = 1 - \frac{v_y}{c},$$  Eq. S4

and coincides with Eq. 1 in the main text. Similarly, for the same condition Eq. S3 reduces to:

$$v_{x'}^o = v_x^o = \frac{c}{1 - \cot(\theta)},$$

which coincides with Eq. 2 in the main text.

### Measurements with a SPAD array

To provide independent corroboration of the trends observed via the scanned-gate method we utilized with our intensified CCD camera, we also, more directly, recorded spatially and temporally resolved data with a different detector, namely a 32 ×32 pixel array of single photon avalanche diodes (SPADs), described in detail in Ref. [17] in the main text, a novel camera with $(56 \pm 1)$-ps temporal resolution. The scattering surface, placed 5 m away was imaged onto the array with a 6-mm focal length lens. The illumination source was the same employed in the experiments described in the main text, however the geometry was different as the condition $\phi = \pi/2 - \theta$ was not satisfied. To provide an example of a more general geometry the angle between the illumination source and the camera direction was set to ~ 55 deg. The screen surface was kept flat but its inclination was changed in order to first have subluminal motion of the plane wave in the direction of the camera ($\theta = 36$ deg, $\phi = 17$ deg) for the data shown in Fig. S2a, and then so as to have a superluminal speed ($\theta = 8$ deg, $\phi = 47$), reflected by Fig. S2b. From the data acquired with the SPAD camera it is possible to extract the perceived velocity of the image for the two cases. For the first case ($\theta = 36$ deg, $\phi = 17$ deg) we found $v_{x,1}^0 = (62 \pm 5)$ cm/ns, while for the second case ($\theta = 8$ deg, $\phi = 47$ deg) we found $v_{x,2}^0 = (-33 \pm 5)$ cm/ns. These values agree qualitatively with predictions obtained from the simple model described above. Indeed, from Eq. 3 it follows that $v_{x,1\,model}^0 = (97 \pm 7)$ cm/ns and $v_{x,2\,model}^0 = (-35 \pm 2)$ cm/ns. The discrepancy between the measured and predicted values is largely due to a less precise characterization of the interaction geometry and to the relatively coarse temporal resolution offered by direct use of the SPAD camera (60-ps time bins and minimum temporal resolution, to be compared to the enhanced resolution we achieved via 10-ps scan steps in our use of the iCCD camera). Nevertheless, the same effects reported in the in the main text and observed with the iCCD are still clearly visible with the direct approach utilized with the compact SPAD camera.

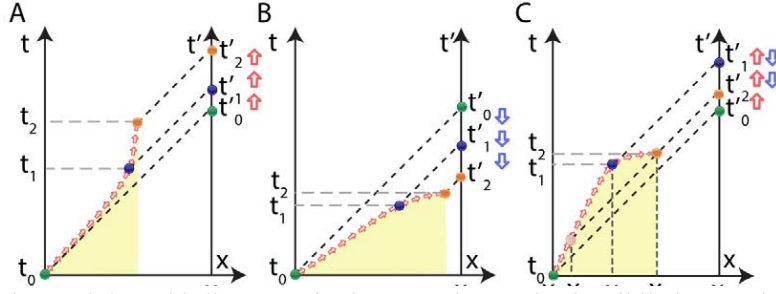

**Figure S3**. Space-time Minkowski diagrams for image pair creation/annihilation: **a** shows an example of a trajectory of a scattering source along a curved surface that, when imaged, maintains the same time-order for all the points. **b** shows a case of a trajectory that results in a complete time reversal. **c** shows a trajectory that features both behaviours and therefore results in the observation of an image pair (pair annihilation).

**Pair creation and annihilation**

We comment here on the origin of the splitting in the images observed for curved scattering surfaces. In Fig. 3a we show the Minkowski space-time diagram that provides geometrical intuition regarding the measured order of the events for a scattering source moving along a subluminal *curved* trajectory in the (*x-t*) space. The relation reported in Eq. S4 for a spatial dependent velocity, maps the time to the measured time:

$$\frac{dt_o}{dt}(x) = 1 - \frac{v(x)}{c} = 1 - \frac{1}{c}\frac{dx}{dt}(x)$$

Eq. S5

If $\forall x, v(x) < c$, the Jacobian in Eq. S5 is always positive, and therefore the relation from $t$ to $t_o$ is invertible (the Jacobian is nonzero) and the time order is maintained. If $\forall x, v(x) > c$, the Jacobian is negative and nonzero, therefore the relation can be inverted and the time order is reversed (see Fig. 3b). However, if $\exists x_0: v(x_0) = c$, then the relation is no longer invertible and becomes multiple valued (non-biunivocal). Figure S3c shows an example of such a situation. At point $x_1$ the local speed passes from subluminal to superluminal and therefore the Jacobian has a zero. The flow of information emanating from spot positions $x_1 < x < x_2$ is mapped onto the observer's time axis coordinates (and is indicated by the procession of blue arrows). We have also mapped the flow of information from spot positions $x_0 < x < x_1$ (indicated by red arrows proceeding from $t'_0$ to $t'_1$). Note that information from spot position $x_c$ reaches the camera at the same instant as information from spot position $x_2$. At this instant, the camera detects both an image of the spot moving between $x_c$ and $x_1$, mapped with a positive Jacobian (therefore preserving the order) *and* the image of the spot moving between $x_1$ and $x_2$, mapped with a negative Jacobian (therefore inverting the time order). These two images appear to the observer to approach one another until "annihilating" where they meet.